\documentclass[12pt,preprint]{aastex}
\usepackage{graphicx}

\usepackage{lineno}
\linenumbers 

\slugcomment{Accepted by ApJ, Draft version 6, \today}

\begin{document}
\renewcommand{\arraystretch}{0.9}

\title{{\it Fermi} LAT observations of the Geminga pulsar}
\author{
A.~A.~Abdo\altaffilmark{1,2}, 
M.~Ackermann\altaffilmark{3}, 
M.~Ajello\altaffilmark{3}, 
L.~Baldini\altaffilmark{4}, 
J.~Ballet\altaffilmark{5}, 
G.~Barbiellini\altaffilmark{6,7}, 
D.~Bastieri\altaffilmark{8,9}, 
B.~M.~Baughman\altaffilmark{10}, 
K.~Bechtol\altaffilmark{3}, 
R.~Bellazzini\altaffilmark{4}
% ...
}
\altaffiltext{1}{Space Science Division, Naval Research Laboratory, Washington, DC 20375, USA}
\altaffiltext{2}{National Research Council Research Associate, National Academy of Sciences, Washington, DC 20001, USA}
\altaffiltext{3}{W. W. Hansen Experimental Physics Laboratory, Kavli Institute for Particle Astrophysics and Cosmology, Department of Physics and SLAC National Accelerator Laboratory, Stanford University, Stanford, CA 94305, USA}
\altaffiltext{4}{Istituto Nazionale di Fisica Nucleare, Sezione di Pisa, I-56127 Pisa, Italy}
\altaffiltext{5}{Laboratoire AIM, CEA-IRFU/CNRS/Universit\'e Paris Diderot, Service d'Astrophysique, CEA Saclay, 91191 Gif sur Yvette, France}
\altaffiltext{6}{Istituto Nazionale di Fisica Nucleare, Sezione di Trieste, I-34127 Trieste, Italy}
\altaffiltext{7}{Dipartimento di Fisica, Universit\`a di Trieste, I-34127 Trieste, Italy}
\altaffiltext{8}{Istituto Nazionale di Fisica Nucleare, Sezione di Padova, I-35131 Padova, Italy}
\altaffiltext{9}{Dipartimento di Fisica ``G. Galilei", Universit\`a di Padova, I-35131 Padova, Italy}
\altaffiltext{10}{Department of Physics, Center for Cosmology and Astro-Particle Physics, The Ohio State University, Columbus, OH 43210, USA}
% ...

\begin{abstract}
%
We report on the \textit{Fermi}-LAT observations of the Geminga pulsar, the second brightest non-variable GeV
source in the $\gamma$-ray sky and the first example of a radio-quiet $\gamma$-ray pulsar. The observations cover
one year, from the launch of the $Fermi$ satellite through 2009 June 15. A data sample of over 60,000 photons
enabled us to build a timing solution based solely on $\gamma$ rays. Timing analysis shows two prominent peaks,
separated by $\Delta \phi$ = 0.497 $\pm$ 0.004 in phase, which narrow with increasing energy. Pulsed
$\gamma$ rays are observed beyond 18 GeV, precluding emission below 2.7 stellar radii because of magnetic
absorption. The phase-averaged spectrum was fitted with a power law with exponential cut-off of spectral index
$\Gamma$ = (1.30 $\pm$ 0.01 $\pm$ 0.04), cut-off energy $E_{0}$ = (2.46 $\pm$ 0.04 $\pm$ 0.17) GeV and an
integral photon flux above 0.1 GeV of (4.14 $\pm$ 0.02 $\pm$ 0.32) $\times$ 10$^{-6}$ cm$^{-2}$ s$^{-1}$. The
first uncertainties are statistical and the second are systematic. The phase-resolved spectroscopy shows a clear
evolution of the spectral parameters, with the spectral index reaching a minimum value just before the leading
peak and the cut-off energy having maxima around the peaks. Phase-resolved spectroscopy reveals that pulsar
emission is present at all rotational phases. The spectral shape, broad pulse profile, and maximum photon energy
favor the outer magnetospheric emission scenarios.
\end{abstract}

\keywords{gamma rays: observations; pulsars: general; pulsars: individual (PSR J0633+1746, Geminga)}

\section{Introduction}
The Geminga pulsar is the second brightest non-variable
GeV $\gamma$-ray source in the sky and the first representative
of a population of radio-quiet $\gamma$-ray pulsars.
Since its discovery as a $\gamma$-ray source by SAS-2, more
than thirty years ago \citep{fichtel75,kniffen75}, Geminga
has been alternatively considered as a unique object or
as the prototype of a population of hidden dead stars.
$Fermi$ has now settled this question with the discovery \citep{abdo09g} of a substantial population of 
potentially radio-quiet pulsars, of which Geminga was indeed the harbinger.\\
Geminga was then observed by the COS B $\gamma$-ray telescope
\citep{bennett77,masnou81}, appearing as 2CG 195+04 in the
second COS B catalog \citep{swanenburg81} and eventually
acquiring the name Geminga \citep{bignami83}.
The X-ray source 1E 0630+178 detected by the \emph{Einstein Observatory}
in the COS B error box \citep{bignami83} was proposed
as a possible counterpart, and subsequently an optical
candidate was found within the \emph{Einstein} error box \citep{bignami87},
which was the bluest object in the field \citep{halpern88,bignami88}.\\
The subsequent ROSAT detection of periodic X-rays from this source
\citep{halpern92} prompted a successful search for periodicity in
high-energy $\gamma$ rays with EGRET \citep{bertsch92}.\\
Geminga has a period of 237 ms and a very stable period
derivative of 1.1 $\times$ 10$^{-14}$ s s$^{-1}$, that
characterize it as a mature pulsar with characteristic age of 3
$\times$ 10$^{5}$ yr and spin-down luminosity $\dot{E}$ = 3.26
$\times$ 10$^{34}$ erg s$^{-1}$.\\
The
determination of the period derivative allowed detection of
$\gamma$-ray pulsations in the previous COS B \citep{bignami92}
and SAS-2 data \citep{mattox92}.
Meanwhile, a high proper motion of 170 mas\//yr for the faint m$_{V}$ = 25.5 optical counterpart was found, confirming the object to be
both underluminous and no more than few hundred pc away
\citep{bignami93}. Using HST, \citet{caraveo96}
obtained a parallax distance for Geminga of 157$^{+59}_{-34}$ pc. A comprehensive review of the history of the identification of
Geminga can be found in \citet{bignami96}.\\
Subsequently, high resolution astrometry with the \emph{Hipparcos}
mission allowed for a 40 mas absolute positioning of Geminga
\citep{caraveo98}. Such accurate positioning, together with the
source proper motion, was used by \citet{mattox98} to improve
the quality of the timing solution of the pulsar.
 Recent parallax and proper motion
measurements confirm the earlier results, yielding a distance of
250$^{+120}_{-62}$ pc
and a proper motion of 178.2 $\pm$ 0.4 mas\//yr \citep{faherty07}.\\
Analysis of EGRET data showed a double peaked light curve with a
peak separation of $\sim$ 0.5 in phase \citep{mayer94,fierro98}.
The Geminga spectrum measured by EGRET was compatible with a
power law with a falloff at $\sim$ 2 GeV, but the limited EGRET
statistics did not allow a measurement of the cut-off energy.
Deep X-ray observations allowed \textit{XMM-Newton} and \textit{Chandra} to map
the neutron star surface as it rotates, bringing into view
different regions contributing different spectral components
\citep{caraveo04,deluca05,jackson05} as well as an arcmin-scale bow-shock feature trailing the
pulsar's motion \citep{caraveo03,deluca06}.
A synchrotron origin of such a non-thermal diffuse X-ray emission trailing the pulsar 
implies the presence of  high-energy electrons ( E$>$ 10$^{14}$ eV, a value close to the upper energy 
limit for pulsar wind electrons in Geminga)  diffusing in a 10 $\mu$G magnetic field.\\
Even though Geminga has been one of the most intensively studied isolated
neutron stars during the last thirty years, it remains of current interest, especially at $\gamma$-ray energies where its
narrow-peaked light curve allows precise timing studies. Thus, it comes as no surprise that Geminga has been a prime target for the $\gamma$-ray instruments currently in operation: AGILE \citep{tavani09} and the Large Area Telescope (LAT) on the $Fermi$ mission \citep{atwood09}. 
Following its
launch, the LAT was confirmed to be an excellent instrument for
pulsar studies, observing the bright Vela pulsar \citep{abdo09a}
and discovering a variety of new $\gamma$-ray pulsars
\citep{abdo09b,abdo09c,abdo09d,abdo09e}, including millisecond
$\gamma$-ray pulsars \citep{abdo09f} and a population of
Geminga-like pulsars detected with blind search techniques \citep{abdo09g}. In this Paper
we present the analysis of the Geminga pulsar based on the
excellent statistics collected during the first year of operations
of the \emph{Fermi} mission.

\section{$\gamma$-ray observations}
The Large Area Telescope (LAT) aboard $Fermi$ is an electron-positron pair conversion telescope sensitive to $\gamma$ rays of energies from 20 MeV to $>$ 300 GeV. The LAT is made of a high-resolution silicon microstrip tracker, a CsI hodoscopic electromagnetic calorimeter and an Anticoincidence detector for charged particles background identification. The full description of the instrument and its performance can be found in \citet{atwood09}.\\ 
The LAT has a large effective area (peaking at $\sim$8000 cm$^{2}$ on axis) and thanks to its field of view ($\sim$ 2.4 sr) covers the entire sky every 2 orbits ($\sim$ 3 h). The LAT point spread function (PSF) strongly depends on both the energy and the conversion point in the tracker, but less on the incidence angle. For 1 GeV normal incidence conversions in the upper section of the tracker the PSF 68$\%$ containment radius is 0.6\degr.\\
The data used in this Paper span roughly the first year of operations after the launch of $Fermi$ on 2008 June
11. The data used for the timing analysis encompass the \emph{Launch and Early Operations} (L$\&$EO), covering
$\sim$ two months after 2008 June 25, when the LAT was operated in pointing and scanning mode for check-out and
calibration purposes, and extend into the first year of nominal operations up to 2009 June 15. For the spectral
analysis we selected only data collected in scanning mode, under nominal configuration, from 2008 August 4 to
2009 June 15. We selected photons in the `diffuse' event class \citep[lowest background contamination,
see][]{atwood09} and we excluded observations when Geminga was viewed at zenith angles $>$ 105\degr~ where
Earth's albedo $\gamma$-rays increase the background contamination. 
We also excluded time intervals when the $15^\circ$ Region Of Interest (ROI) intersects the Earth's albedo region.
\section{Timing Geminga using $\gamma$ rays}\label{sec:timing}
\begin{figure}[ht]
\begin{center}
\includegraphics[width=0.95\columnwidth]{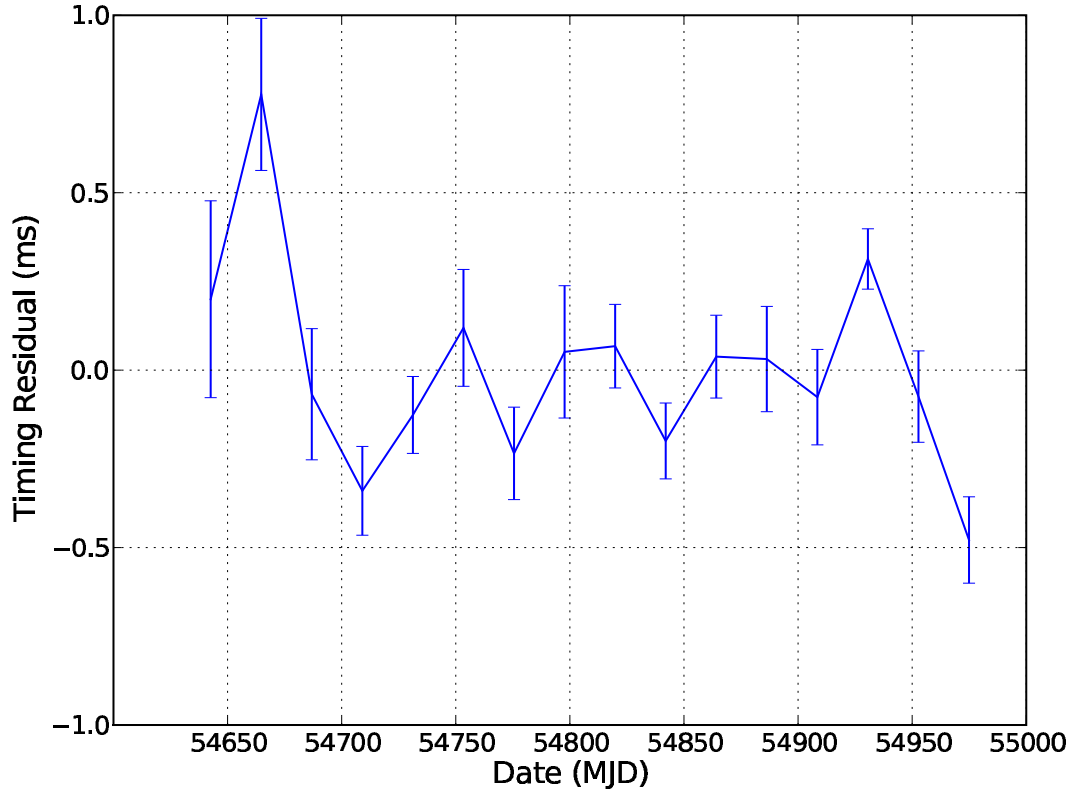}
\caption{Timing residuals of Geminga derived from the model built using the TOAs of the $\gamma$ rays detected by the LAT (See Section \ref{sec:timing} for details).}\label{fig:gemresrms}
\end{center}
\end{figure}

Since the end of the EGRET mission, the Geminga timing ephemeris has been maintained using occasional observations with \textit{XMM-Newton} (Jackson \& Halpern 2005; J. Halpern, private communication). While AGILE relied on such X-ray ephemerides \citep{pellizzoni09}, LAT densely-sampled, high-precision timing observations yielded an independent timing solution. In fact, the LAT timing is derived from a GPS clock on the spacecraft and times of arrival of $\gamma$ rays are recorded with an accuracy significantly better than 1 $\mu$s \citep{abdo09h}. We have constructed a timing solution for Geminga using the \textit{Fermi} LAT data, exclusively. For this analysis, we assumed a constant location for the Geminga pulsar calculated at the center of the time span of the LAT data set (MJD 54800) using the position reported by \citet{caraveo98} and updated according to the source proper motion \citep{faherty07}.\\ 
We determined an initial, approximate, ephemeris using an epoch-folding search.  
We then measured pulse times-of-arrival (TOAs) by first converting the photon event times to a reference 
point at the geocenter using the \textit{Fermi} 
science tool\footnote{http://fermi.gsfc.nasa.gov/ssc/data/analysis/scitools/overview.html}
\emph{gtbary}, then computing a pulse profile using phases generated using \textsc{TEMPO2} \citep{hobbs06} 
in its predictive mode. 
The timing accuracy of \emph{gtbary} was demonstrated in \citet{smith08}.
This was done with $\sim22$ day segments of data. 
TOAs were determined from each segment using a Fourier-domain cross correlation with a high signal-to-noise template profile. We obtained 16 TOAs in this way from 2008 June 25 to 2009 June 15. We fit these TOAs, again using \textsc{TEMPO2}, to a model with only absolute phase, frequency and frequency first derivative as free parameters.  The residuals to the model have an RMS of 251\,$\mu$s, as shown in Figure \ref{fig:gemresrms}, and the model parameters are listed in Table~\ref{Tbl:Ephems}.
The epoch of phase 0.0 given in Table~\ref{Tbl:Ephems} is defined so that the phase of the first component of the Fourier transform of the light curve has 0 phase.
However, in order to assign a smaller phase to the leading peak, we introduced an additional phase shift of 0.5 to the timing solution in Table~\ref{Tbl:Ephems}. Thus, in the light curve shown in Figure \ref{fig:gemlc}, the epoch of phase 0.0 is the barycentric arrival time MJD(TDB) corresponding to phase 0.5.\\ 
\begin{table}[!ht]
\caption{$Fermi$-LAT Ephemeris for Geminga}
\label{Tbl:Ephems}
\centering
\begin{tabular}{lc}
\hline
\hline
Parameter & Value \\
\hline
Epoch of position (MJD)   &  54800 \\
R.A. (J2000)              &  6:33:54.289 \\
Dec. (J2000)              &  +17:46:14.38 \\
\hline
\\
\hline
Epoch of ephemeris T$_0$ (MJD) &  54800 \\
Range of valid dates (MJD)     &  54642 -- 54975 \\
Frequency f (s$^{-1}$)               &  4.21756706493(4) \\
Freq. derivative $\dot{f}$ ($\times 10^{-13}$ s$^{-2}$) & -1.95250(9) \\
Freq. 2nd derivative $\ddot{f}$ (s$^{-3}$) & 0 \\
Epoch of Phase 0.0 (MJD(TDB)) & 54819.843013078(3) \\
Time Units & TDB\\
\hline
\end{tabular}
\end{table}
\section{Results}
\subsection{Light curves}\label{sec:an_pulse}
The strong energy dependence of the PSF imposes energy-dependent regions of interest (ROI) that optimize the signal to noise ratio. 
Following a procedure similar to that used for the \textit{Fermi}-LAT pulsar catalogue paper \citep{abdo09l}, 
to study the pulse profiles we selected photons within an angle $\theta$ $<$ max[1.6-3$\log_{10}$(E$_{GeV}$),1.3] degrees from Geminga. 
Such selection provides clean light curves by limiting acceptance of the softer Galactic background.

\begin{figure}[ht]
\begin{center}
\includegraphics[width=1.\columnwidth]{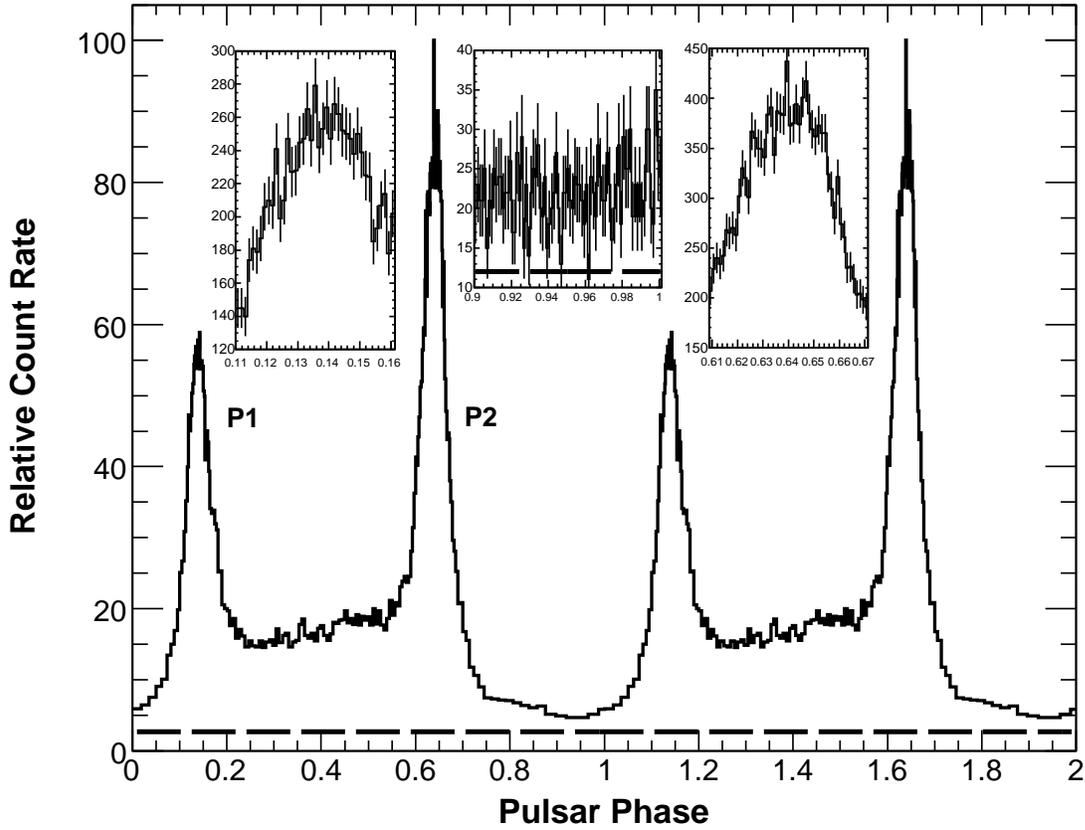}
\caption{Geminga light curve above 0.1 GeV using an energy-dependent ROI, shown over two pulse periods. The count rate is shown in variable-width bins, each one containing 400 counts per bin and normalized to 100. Insets show the phase intervals centered on the two peaks and on the ''second interpeak'' region ($\phi$ = 0.9--1.0), binned to 0.00125 in phase. The dashed line represents the contribution of the diffuse background estimated by selecting photons in this ''second interpeak'' interval in an annulus around the source.}\label{fig:gemlc}
\end{center}
\end{figure}

\begin{figure}[ht]
\begin{center}
\includegraphics[width=0.95\columnwidth]{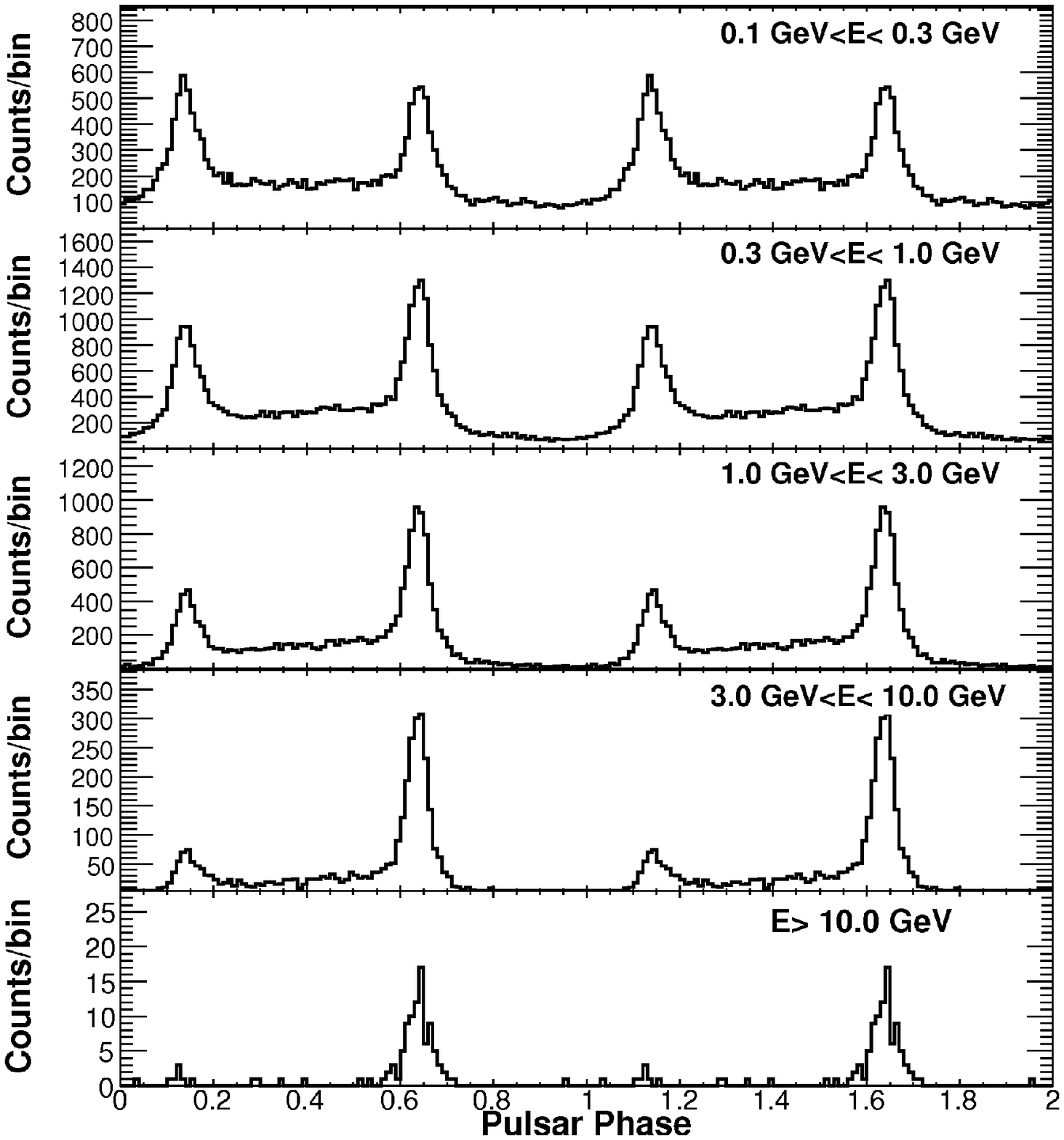}
\caption{Geminga light curves in five energy ranges (0.1--0.3 GeV, 0.3--1 GeV, 1--3 GeV, 3--10 GeV, $>$ 10 GeV). Each light curve is shown over two pulse periods and contains 100 bins/period.}\label{fig:geminga_edep}
\end{center}
\end{figure}

\begin{figure}[ht]
\begin{center}
\includegraphics[width=0.95\columnwidth]{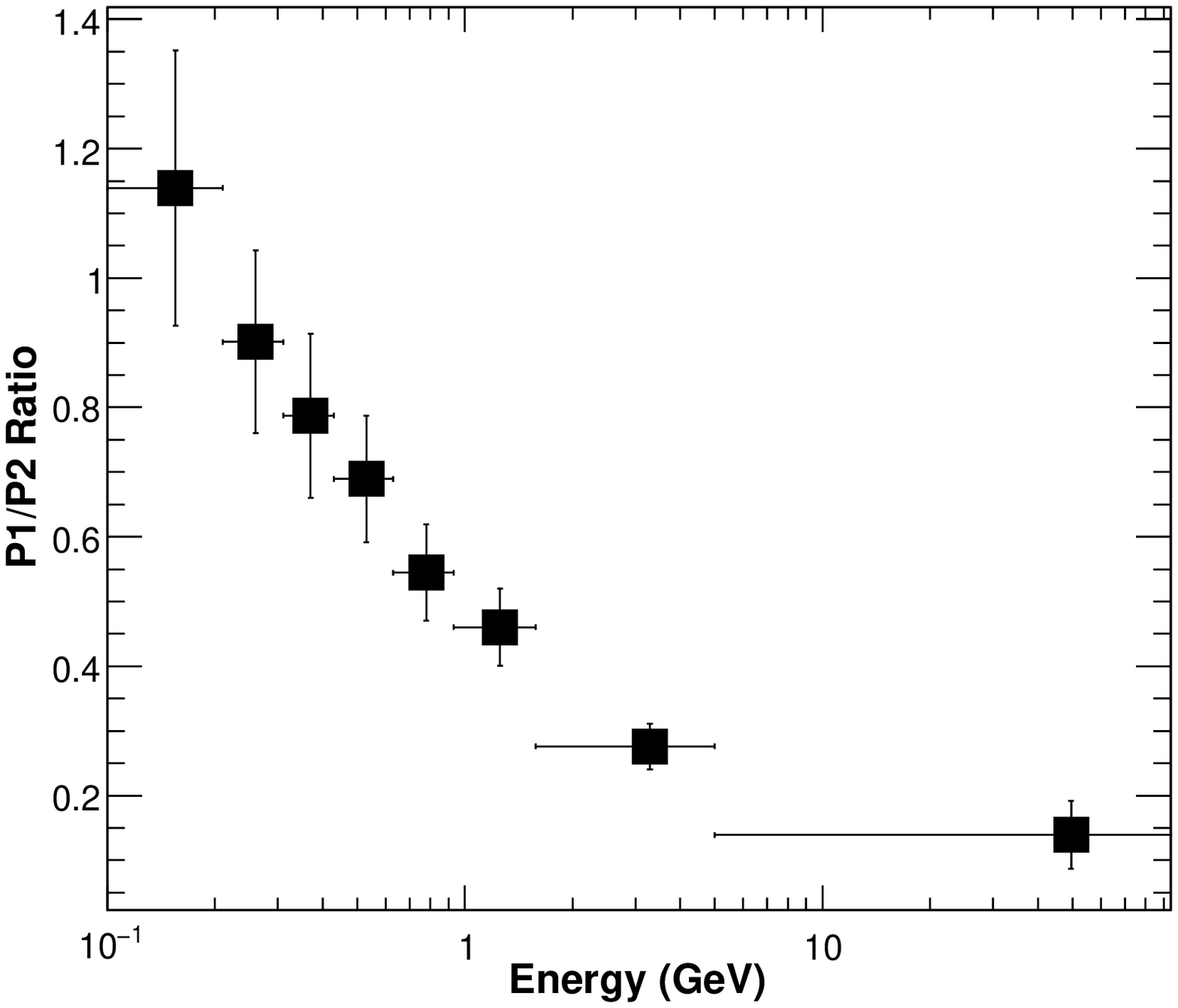}
\caption{Evolution of the ratio P1/P2 with energy, plotted in variable-width energy bins, each one containing 10000 events.}\label{fig:geminga_p1p2}
\end{center}
\end{figure}

\begin{figure}[ht]
\begin{center}
\includegraphics[width=0.95\columnwidth]{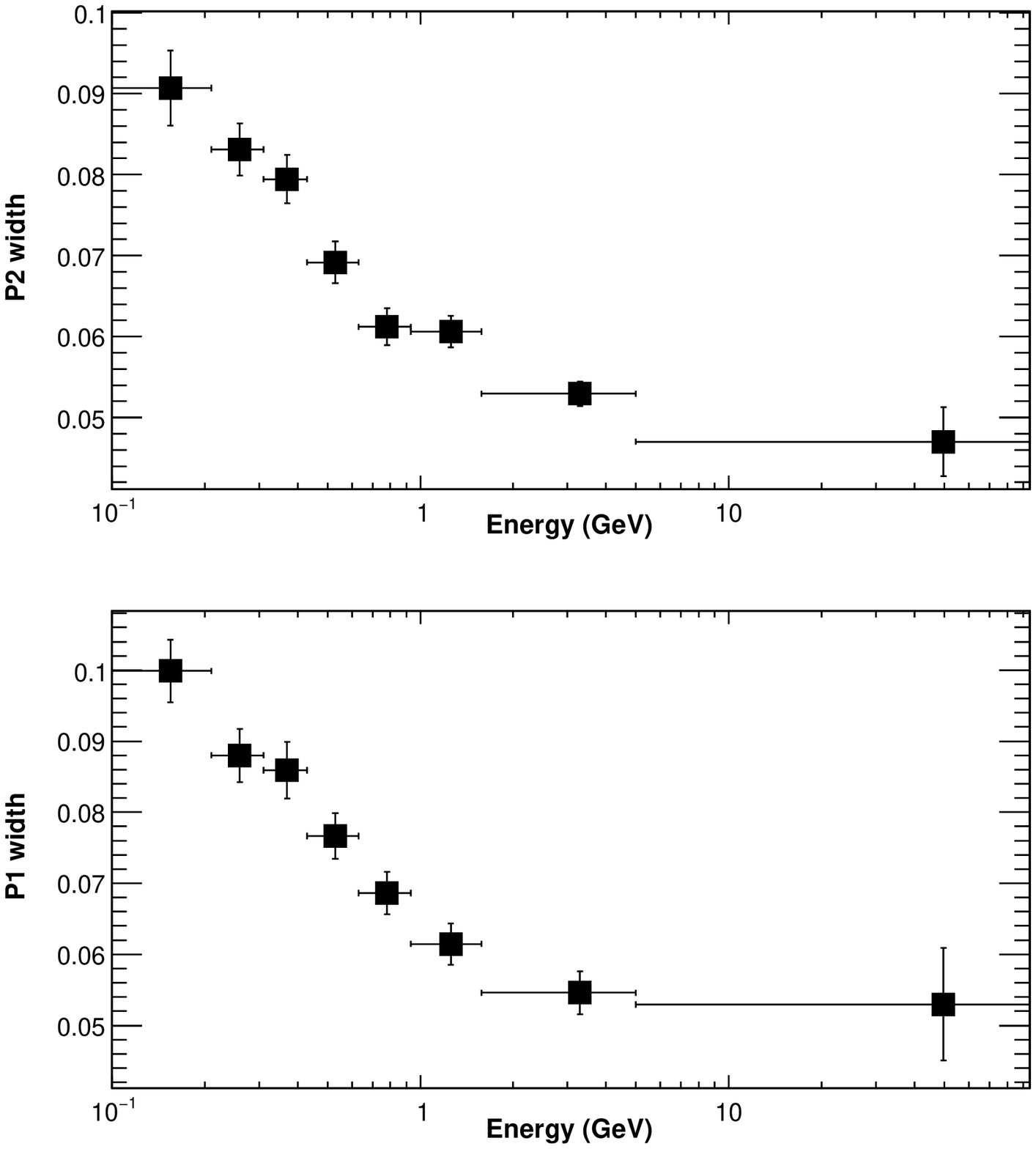}
\caption{Evolution with energy of the FWHM of P1 (\textit{bottom}) and P2 (\textit{top}), plotted in variable-width energy bins, each one containing 10,000 events. Both peaks narrow at increasing energies.}\label{fig:geminga_pwidth}
\end{center}
\end{figure}

We used the \textit{Fermi} tool \textit{gtpphase} to correct photon arrival times to the Solar System barycenter using 
the JPL~DE405 Solar System ephemeris \citep{standish98} and to assign a rotational phase to each photon using the timing solution described in Section \ref{sec:timing}.\\
Figure \ref{fig:gemlc} shows the light curve of Geminga above 0.1 GeV obtained with the energy-dependent cut. In order to better show the fine structure, we plot the pulse profile using variable-width phase bins, each one containing 400 events. 
The photon flux in each phase interval thus has a 1$\sigma$ Poisson statistical error of 5$\%$. 
The dashed line represents the contribution of the diffuse background, estimated by selecting photons in the 
phase interval $\phi$ = 0.9--1.0 from an annular region between 2\degr and 3\degr~from the source rescaled for 
the solid angle and also taking into account the energy-dependent selection adopted. The light curve contains
61219 $\pm$ 284 pulsed photons and 9821 $\pm$ 99 background photons.\\
The pulse profile shows two clear peaks at $\phi$ = 0.141 $\pm$ 0.002 (P1) and $\phi$ = 0.638 $\pm$ 0.003 (P2). 
In order to reveal possible asymmetries in the peaks, we started by fitting the sharp peaks with two half-Lorentzian profiles with different widths for the trailing and the leading edge. We have chosen this function because it has a simple parameterization and appear to fit well the pulse profile of the gamma-ray light curves. We found that Geminga peaks show no asymmetries, and P1 is broader (FWHM of 0.072 $\pm$ 0.002) than P2 (FWHM 0.061 $\pm$ 0.001). We also checked if the peaks can be better fitted by a Gaussian profile, finding comparable results (P1 FWHM  of 0.071 $\pm$ 0.002) and (P1 FWHM  of 0.063 $\pm$ 0.001), though we cannot distinguish between a Lorentz or Gaussian profile. The smallest features in the pulse profile appear on a scale of 260 $\mu$s, presumably artifacts of the timing model residuals. Figure \ref{fig:gemlc} also contains insets (binned to 0.00125 in phase) centered on the two peaks and on the phase interval $\phi$ = 0.9--1.0. This off-peak, or ``second interpeak'', region  contains 789 $\pm$ 28 pulsed photons above the estimated background ($\sim$ 1.3 $\times$ 10$^{-2}$ of the pulsed flux). This corresponds to a signal-to-noise ratio of 19$\sigma$, indicating that the pulsar emission extends also in the off-peak, as will be investigated further in Section 4.3.\\
Figure \ref{fig:geminga_edep} shows the pulse profile in five energy ranges (0.1--0.3 GeV, 0.3--1 GeV, 1--3 GeV, 3--10 GeV, $>$ 10 GeV). There is a clear evolution of the light curve shape with energy: P1 becomes weaker with increasing energy, while P2 
is still detectable at high energies. 
Significant pulsations from P2 are detectable at energies beyond $\epsilon_{max} \sim 18$ GeV,
chosen as the maximum energy beyond which a $\chi^{2}$ periodicity test still attains $6\sigma$ significance. 
We detect 16 photons above 18 GeV, not necessarly coming from the pulsar itself.
No particular features appear at high energies in the bridge region between P1 and P2 (''first interpeak'').\\
Figure \ref{fig:geminga_p1p2} shows the evolution of the P1/P2 ratio as a function of energy, plotted using variable-width energy bins.
The curve depends very weakly on the bin choice, Figure 4 was made using 10000 events per bin. A clear decreasing trend is visible, as observed in Crab, Vela and PSR B1951+32 $\gamma$-ray pulsars by EGRET \citep{thompson04} and now confirmed for the Vela \citep{abdo09a} and the Crab pulsars \citep{abdo10c} by \emph{Fermi} LAT.
Adopting the same variable-width energy bins we fit the peaks in each energy range with a Lorentz function to determine the peak center and width. Figure \ref{fig:geminga_pwidth} shows the energy evolution of the FWHM of P1 and P2: both peaks narrow with increasing energy. The decreasing trend in pulse width of P1 and P2 is nearly identical. P1 has a FWHM decreasing from $\delta\phi$ = 0.098 $\pm$ 0.004 to $\delta\phi$ = 0.053 $\pm$ 0.008, while FWHM of P2 changes from $\delta\phi$ = 0.092 $\pm$ 0.004 to $\delta\phi$ = 0.044 $\pm$ 0.004 at energies greater than 3 GeV. The decrease in width with energy does not depend on the shape used to fit the peaks. Figure 8 was made using the Lorenztian fits, preferred in general because sensitive to asymmetric pulses. While the ``first interpeak'' emission is significantly detected up to 10 GeV, emission in the ``second interpeak'' region (between 0.9 and 1.0), not detected before, is clearly present at low energies but vanishes above $\sim$ 2 GeV.\\
\subsection{Energy Spectrum}\label{sec:an_spec}
\begin{figure}[ht]
\begin{center}
\includegraphics[width=0.95\columnwidth]{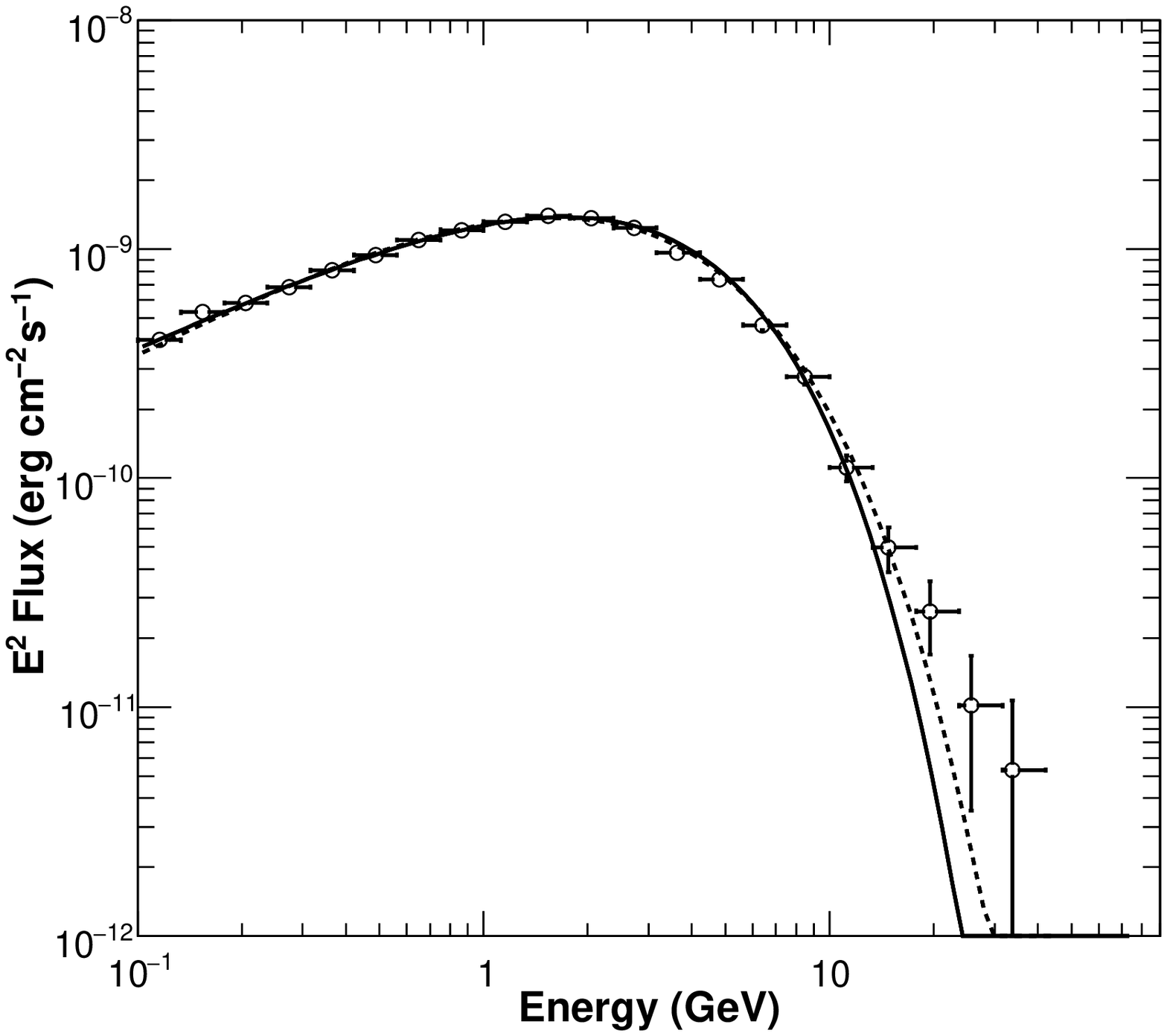}
\caption{Phase-averaged Spectral Energy Distribution (SED) of the Geminga pulsar. The solid line represents the best fit power law with exponential cutoff (i.e b=1), while the dashed one represents the best fit power law with exponential cutoff with free exponential index (in this case the result is b=0.81). The LAT spectral points (open circles) are obtained using the maximum likelihood method described in Section 4.2}\label{fig:geminga_phaseave}
\end{center}
\end{figure}

Spectral analysis was performed using the maximum-likelihood estimator \emph{gtlike} included in the standard \emph{Fermi} Science Tools provided by the FSSC. The fit was performed using a region of the sky with a radius of 15\degr~around the pulsar position selecting energies between 0.1 and 100 GeV.\\ 
We included in the fit a model accounting for the diffuse emission as well as for the nearby $\gamma$-ray sources. We modeled the diffuse foreground, including Galactic interstellar emission, extragalactic $\gamma$-ray emission and residual CR background, using the  models\footnote{http://fermi.gsfc.nasa.gov/ssc/data/access/lat/BackgroundModels.html} gll$\_$iem$\_$v02 for the Galactic part and isotropic$\_$iem$\_$v02 for the isotropic one.\\
In the fit procedure we fixed the spectral parameters of all the sources between 15$^\circ$ and 20$^\circ$ from Geminga, and left free the normalization factor of all the sources within 15$^\circ$. All the non-pulsar sources have been modeled with a power law as reported in the $Fermi$ Bright Source List \citep{abdo09l}, while all the pulsars have been described by a power law with exponential cut-off according to the data reported in the $Fermi$-LAT pulsar catalogue \citep{abdo09l}.\\
We integrated the phase-averaged spectrum to obtain the energy flux. The unbinned $gtlike$ fit is described by a power law with exponential cut-off in the form:
\begin{equation}
\frac{dN}{dE} = N_{0} E^{-\Gamma} \exp \left(- \frac{E}{E_{0}}\right)  \mbox{cm$^{-2}$s$^{-1}$GeV$^{-1}$}
\label{eq_plec}
\end{equation}
where $N_{0}$ = (1.189 $\pm$ 0.013 $\pm$ 0.070) $\times$ 10$^{-9}$ cm$^{-2}$ s$^{-1}$ GeV$^{-1}$, $\Gamma$ = (1.30 $\pm$ 0.01 $\pm$ 0.04) and $E_{0}$ = (2.46 $\pm$ 0.04 $\pm$ 0.17) GeV. The first uncertainties are statistical values for the fit parameters, while the second ones are systematic uncertainties. Systematics are mainly based on uncertainties on the LAT effective area derived from the on-orbit estimations, and are of $\leq$ 5$\%$ near 1 GeV, 10$\%$ below 0.1 GeV and 20$\%$ above 10 GeV. We therefore propagate these uncertainties using modified effective areas bracketing the nominal ones (P6$\_$v3$\_$diffuse).\\
For this fit over the range 0.1 -- 100 GeV we obtained an integral photon flux of (4.14 $\pm$ 0.02 $\pm$ 0.32) $\times$ 10$^{-6}$ cm$^{-2}$ s$^{-1}$ and a corresponding energy flux of (4.11 $\pm$ 0.02 $\pm$ 0.27) $\times$ 10$^{-9}$ erg cm$^{-2}$ s$^{-1}$.\\
We studied alternative spectral shapes beginning with the cut-off function $\exp[-(E/E_{0})^{b}]$. 
The 46 gamma-ray pulsars discussed in \citet{abdo10a} are generally well-described by a simple exponential cutoff, $b$ = 1, a shape predicted by outer magnetosphere emission models (see the Discussion, below). Models where gamma-ray emission occurs closer to the neutron star can have sharper ``super-exponential'' cutoffs, e.g. $b$ = 2. Leaving free the exponential index $b$ we obtained $N_{0}$=  (1.59 $\pm$ 0.13 $\pm$ 0.09) $\times$ 10$^{-9}$ cm$^{-2}$ s$^{-1}$ GeV$^{-1}$, $\Gamma$ = (1.18 $\pm$ 0.03 $\pm$ 0.04), $E_{0}$=1.58 $\pm$ 0.19 $\pm$ 0.11) GeV and $b$=(0.81 $\pm$ 0.03 $\pm$ 0.06). As previously reported for the analysis of Vela pulsar \citep{abdo10b}, $b$ $<$ 1 can be interpreted  by a blend of $b$ = 1 spectra with different cutoff energies. Figure \ref{fig:geminga_phaseave} shows the results of the phase-averaged spectrum in case of $b$ free (dashed line) and $b$ fixed to 1 (solid line).  Using the likelihood ratio test we found that the hypothesis of $b$=2 can be excluded since the likelihood of this fit being a good representation of the data is much greater than for a power-law fit (logarithm of the likelihood ratio being 396) . We have also tried different spectral shapes, like a broken power law, but the fit quality does not improve (the logarithm of the likelihood ratio is 212).
\subsection{Phase-resolved analysis}
\begin{figure}[ht]
\begin{center}
\includegraphics[width=0.95\columnwidth]{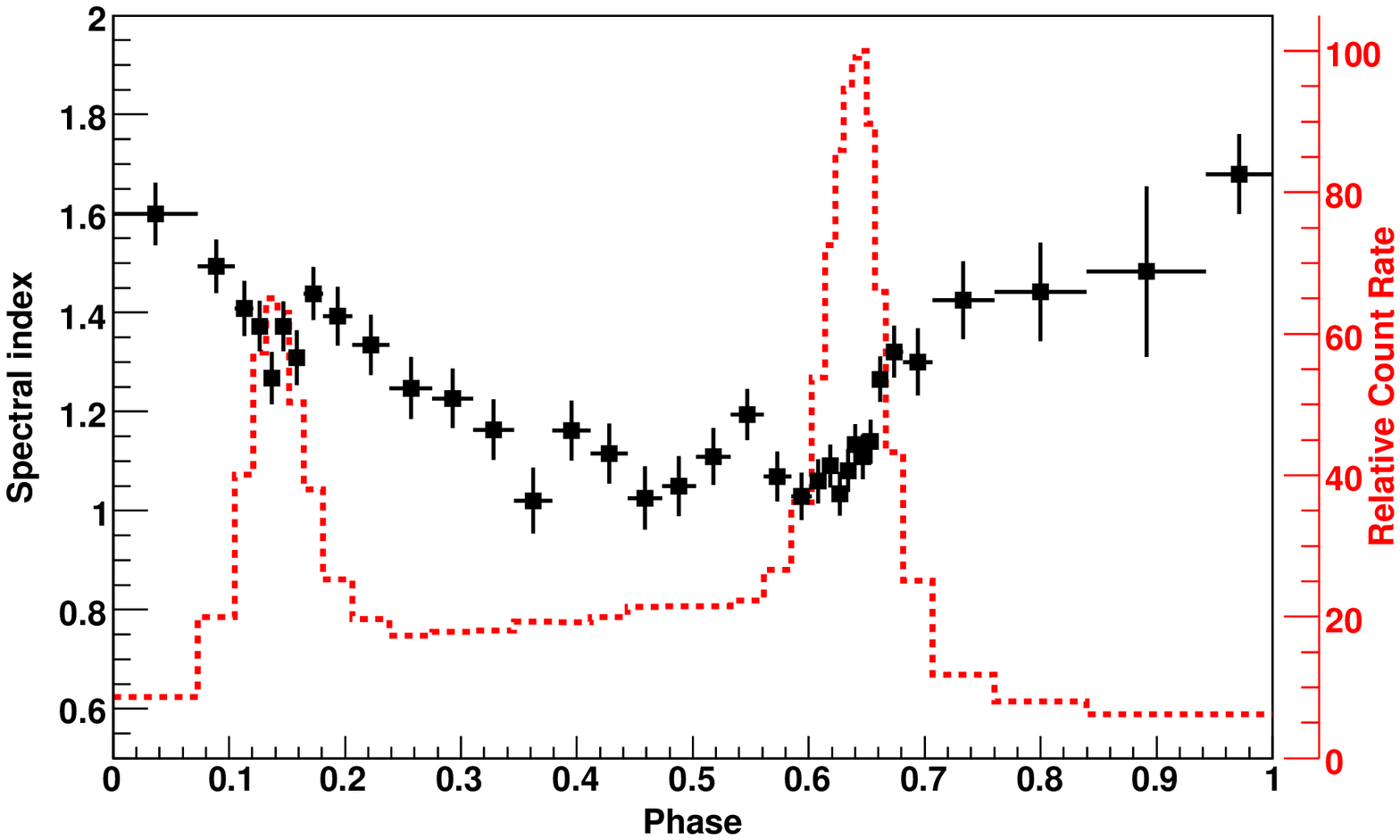}
\includegraphics[width=0.95\columnwidth]{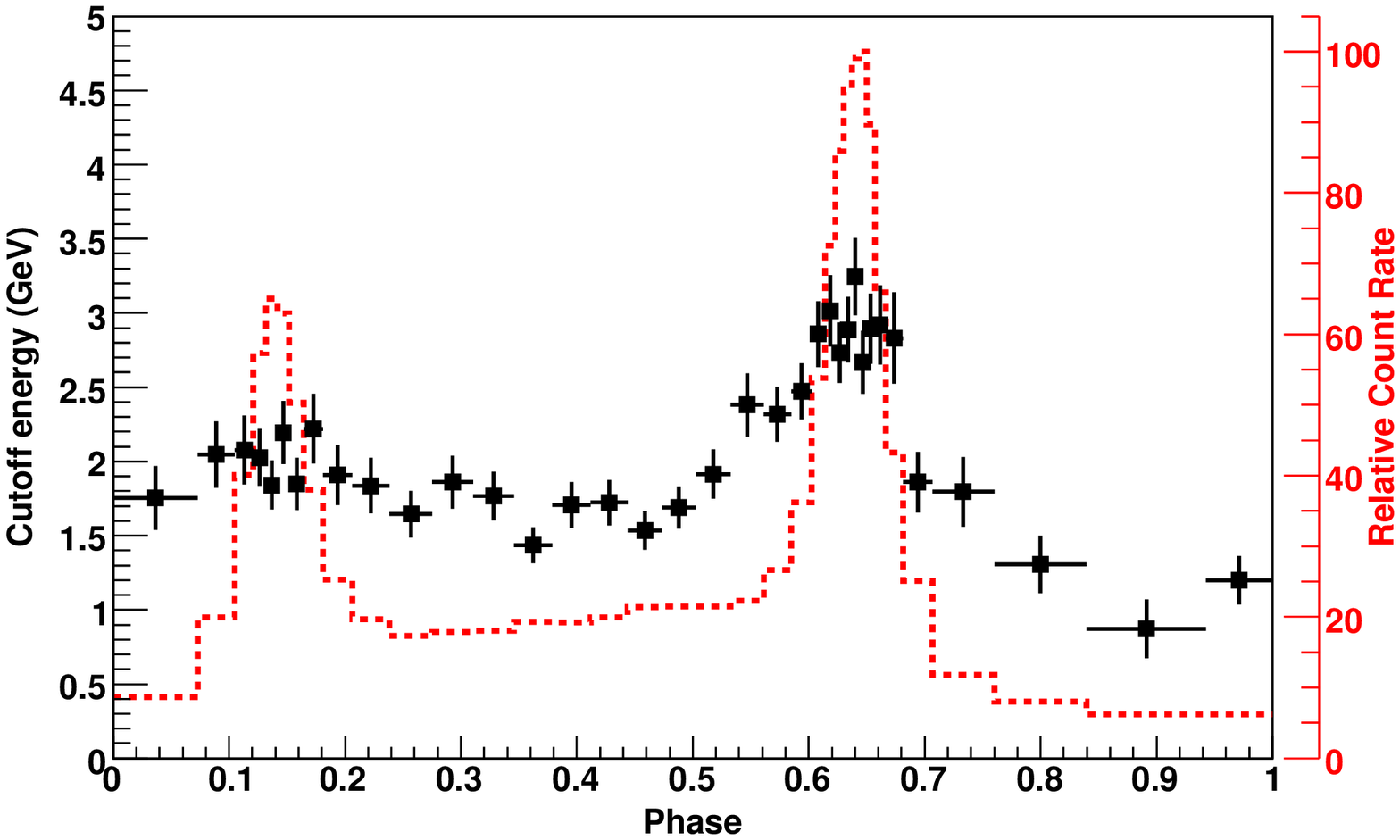}
\caption{Phase evolution of the spectral index (\emph{top}) and energy cut-off (\emph{bottom}) above 0.1 GeV as the function of the pulse phase, divided in phase bins each containing 2000 photons. Vertical bars indicate the combined statistical and systematic uncertainties. For each phase interval (defined in Table \ref{tab:phres} in the Appendix) a power law with exponential cut-off has been assumed. The dashed histogram represents the $Fermi$-LAT light curve above 0.1 GeV in variable-width phase bins of 2000 photons/bin.}\label{fig:ph_evol}
\end{center}
\end{figure}

\begin{figure}[ht]
\begin{center}
\includegraphics[width=0.95\columnwidth]{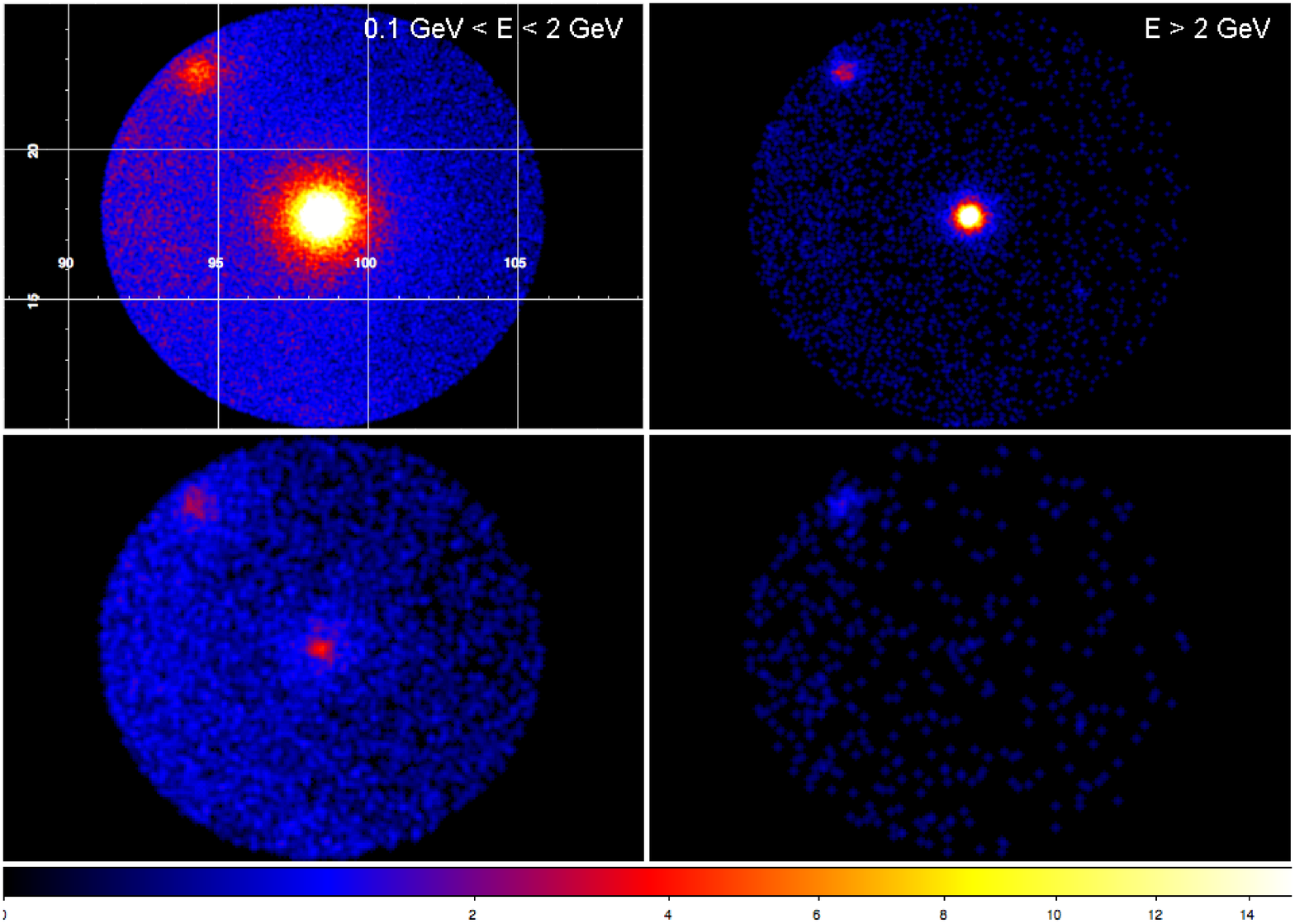}
\caption{Maps representing the phase interval ($\phi$=0.0 -- 0.9, \textit{top row}) compared to the second interpeak ($\phi$=0.9-1.0, \textit{bottom row}), in the 2 energy bands 0.1--2 GeV and $>$ 2 GeV. Each map represents the photons within 7$^\circ$ from Geminga, binned in pixels of 0.045$^{\circ}$ (\textit{top row}) and 0.09$^{\circ}$ (\textit{bottom row}),smoothed with a gaussian filter with a radius of 2 pixels. In the upper left panel we reported the Right Ascension in horizontal axis and the Declination in the vertical axis. Bottom row shows that the offpeak point source image is visible at low energies but vanishes at E $>$ 2 GeV due to the spectral cut-off.}\label{fig:ph_map}
\end{center}
\end{figure}

We divided the pulse profile in variable-width phase bins, each one containing 2000 photons according to the energy-dependent cut defined in Section \ref{sec:an_pulse}.This choice of binning provides a reasonable compromise between the number of photons needed to perform a spectral fit and  the length of the phase intervals, that should be short enought to sample fine details on the lightcurve, while remaining confortably larger than the rms of the timing solution (Sec. 3). We have performed a maximum likelihood spectral analysis, similar to the phase-averaged one, in each phase bin assuming a power law with exponential cut-off describing the spectral shape.  Using the likelihood ratio test we checked that we can reject the power law at a significance level greater than 5 sigma in each phase interval. Following the results on phase-averaged analysis of Geminga, we have modeled the spectrum in each phase interval with a power law with exponential cutoff. Such a model yields a robust fit with a logarithm of the likelihood ratio greater than 430 in each phase interval. Figure \ref{fig:ph_evol} (below) shows the evolution of the spectral parameters across Geminga's rotational phase. In particular, the energy cutoff trend provides a good estimate of the high energy emission variation as a function of the pulsar phase.
Table \ref{tab:phres} summarizes the results of the spectral fit in each phase bin. In this case we have fixed all the spectral parameters of all the nearby $\gamma$-ray sources and of the two diffuse backgrounds to the values obtained in the phase averaged analysis, rescaled for the phase bin width.\\
To obtain \textit{Fermi}-LAT spectral points we divided our sample into logarithmically-spaced energy bins (4 bins per decade
starting from 100 MeV) and then applied the maximum likelihood method in each bin. For each energy bin we have used a model with all the nearby sources as well as Geminga described by power law with fixed spectral index. We have considered only energy bins in which the source significance was greater than  $3\sigma$. 
From the fit results we then evaluated the integral flux in each energy bin. 
This method does not take energy dispersion into account and correlations among the energy bins. 
To obtain the points of the Spectral Energy Distributions (SEDs) we multiplied each bin by the mean energy value of the bin taking into account the spectral function obtained by the overall fit.
Figures \ref{fig:app_phres1} to \ref{fig:app_phres4} in the Appendix show the SEDs obtained in each phase interval. The fluxes in Y-axis are not normalized to the phase bin width, whereas in Table \ref{tab:phres} of the Appendix the fluxes are normalized.
Figure \ref{fig:ph_evol} shows the phase evolution of the spectral index and cut-off energy, respectively. The spectral index reaches a local minimum around P1 ($\phi$ $\sim$ 0.14 -- 0.15) and, after a sudden increase, 
begins to decrease again in the ``first interpeak'' region, reaching a minimum of $\Gamma$ $\sim$ 1.1 around the leading edge of P2 ($\phi$ $\sim$ 0.60 -- 0.61). 
It then starts to rise again in the phase interval from P2 to the ``second interpeak'' region ($\phi$ = 0.9 -- 1.0).\\ 
The cut-off energy evolves quite differently as a function of the rotational phase. It closely follows the pulse profile, thus confirming the observations performed by EGRET \citep{fierro98}, which unveiled a correlation between hardness ratio and pulse profile. As shown in EGRET data and recently confirmed by AGILE \citep{pellizzoni09}, the hardest component is P2: our phase-resolved scan points to a cut-off around 3 GeV and a spectral index of $\sim$ 1.0 that become softer through the peak. P1 appears to be softer, with a cut-off energy slightly greater than 2 GeV and a spectral index $\Gamma$ $\sim$ 1.2.\\ 
The phase-resolved spectra show that Geminga's emission in the bridge (or ``first interpeak'') phase interval ($\phi$ = 0.2
-- 0.52) is quite different from the Crab \citep{abdo10c} or Vela pulsars \citep{fierro98,abdo09a}. For the Crab pulsar the bridge emission shows no evolution and drops to an intensity level comparable to the off pulse emission, while for the Vela pulsar it varies substantially but is always seen at high energies. The ``first interpeak'' of Geminga, instead, becomes harder and remains quite strong at high energies, as can be also seen in Figure \ref{fig:geminga_edep}. Another difference with respect to the Vela pulsar is that Geminga does not have a third peak like the one observed at GeV energies in the Vela pulsar \citep{abdo09a}.\\
The analysis of the ``second interpeak'' region around $\phi$ = 0.9 -- 1.0 shows significant emission up to $\sim$ 2 GeV
(Figure \ref{fig:geminga_edep}). Moreover the spectrum in this phase interval has been fit with a power law with exponential cut-off, obtaining a spectral index $\Gamma$ = (1.48 $\pm$ 0.17) and $E_{0}$ = (0.87 $\pm$ 0.19) GeV, with systematic uncertainties in agreement with those evaluated in the phase averaged analysis. A pure power law fit can be rejected with a $\sim$8$\sigma$ confidence level, thus confirming the presence of the cut-off. The presence of the ``second interpeak'' component is also visible in the maps of
Figure \ref{fig:ph_map}, where the emission in this phase region is not visible at high energies, as expected owing to 
the spectral cut-off.\\
Analyzing the phase evolution of the spectral parameters in Figure \ref{fig:ph_evol} it seems that no abrupt changes occur in this phase interval and that this emission may be related to the wings of the peaks. This fact, together with the newly detected off-peak emission, favors a pulsar origin of such ``second interpeak'' emission, rather than an origin in a surrounding region. The detection of off-peak emission, rendered possible by the outstanding $Fermi$ statistics, is a novelty of Geminga's high energy behaviour.
\section{Discussion}
\subsection{Light curves and beam geometry}
The unprecedented photon statistics collected by $Fermi$ LAT allows for tighter observational constraints on emission models. The absence of radio emission characterizing Geminga clearly favors models where the high energy emission occurs in the outer magnetosphere of the pulsar.\\ 
Polar Cap (PC) models, where high energy emission is located near the neutron star surface \citep{daugherty96}, are unlikely to explain the Geminga pulsar, since the line of sight is necessarily close to the magnetic axis for such models where one expects to see radio emission.\\
The current evidence against low-altitude emission in $\gamma$-ray pulsars \citep{abdo09l} can also be supplemented by constraints on a separate physical origin. In PC models, $\gamma$ rays created near the neutron star surface interact with the high magnetic fields of the pulsar, producing sharp cut-offs in the few to $\sim$ 10 GeV energy regime. 
Moreover, the maximum observed energy of the pulsed photons observed must lie below the $\gamma$--B pair production mechanism threshold, providing a lower bound to the altitude of the $\gamma$-ray emission. According to \citet{baring04}, the lower limit for the altitude of the production region $r$ could be estimated taking advantage of the maximum energy detected for pulsed photons $\epsilon_{max}$ as $r\ge\left(\epsilon_{max}B_{12}/1.76GeV\right)^{\frac{2}{7}}P^{-\frac{1}{7}}R_{*}$ where $P$ is the spin period, $R_{*}$ is the stellar radius and 
$B_{12}$ is the surface magnetic field in units of $10^{12}G$. 
For pulsed photons of $\epsilon_{max}\sim18$ GeV, we obtain  $r_{min}\ge2.7R_{*}$, a value clearly precluding emission very near the stellar surface, adding to the advocacy for a slot gap or outer gap acceleration locale for the emission in this pulsar.\\
Outer Gap (OG) models \citep{cheng86,romani96,zhang01}, where the high energy emission  extends between the null charge surface and the light cylinder, the two-pole caustic  (TPC) models \citep{dyks03} associated with slot gap (SG) \citep{muslimov04}, where the emission is located along the last open field lines between the neutron star surface and the light cylinder, or a striped wind model \citep{petri09}, where the emission originates outside the light cylinder, could produce the observed light curve and spectrum.
Nevertheless, the observed peak separation of 0.5 is unlikely for a middle aged pulsar like Geminga in the OG model, 
if it is true that emission moves to field lines closer to the magnetic axis as pulsars age.
For the OG model this drift leads to $<0.5$ peak separations. For TPC models $0.5$ peak separation can occur
in spite of this shift, that is, for all ages and spin-down luminosities.\\
Following the Atlas of $\gamma$-ray light curves compiled by \citet{watters09}, we can use Geminga's light curve to estimate, for each model, the star's emission parameters, namely the Earth viewing angle $\zeta_{E}$ with respect to the neutron star spin axis, and the inclination angle $\alpha$ between the star's magnetic and rotation axes.
\begin{table}[ht]
\begin{center}
\begin{tabular}{lcccc}
\hline
\hline
Model & $\alpha$ & $\zeta_{E}$ & $f_\Omega$ \\
\hline
TPC & $30-80,90$ & $90,55-80$ & $0.7-0.9,0.6-0.8$ \\
OG & $10-25$ & $85$ & $0.1-0.15$ \\
\hline
\end{tabular}
\caption{Earth viewing angles $\zeta_{E}$, inclination angles $\alpha$ and beaming factor $f_{\Omega}$ for Geminga, as predicted by \citet{watters09} for Outer Gap (OG) and Two Pole Caustics (TPC) models.}
\label{table_watters}
\end{center}
\end{table}
Table \ref{table_watters} summarizes the observed parameters and gives the estimated beaming correction factor 
$f_{\Omega}(\alpha,\zeta_{E})$, which is model-sensitive. It is given by \citep{watters09}:
\begin{equation}\label{eq:fbeam}
f_{\Omega}(\alpha,\zeta_{E}) = \frac{\int F_{\gamma}(\alpha;\zeta,\phi)\sin(\zeta)d\zeta d\phi}{2 \int F_{\gamma}(\alpha;\zeta_{E},\phi) d\phi}
\end{equation}
where $F_{\gamma}(\alpha;\zeta,\phi)$ is the radiated flux as a function of the viewing angle $\zeta$ and the pulsar phase $\phi$. In this equation, the numerator is the total emission over the full sky, and the denominator is the expected phase-averaged flux for the light curve seen from Earth.\\ 
The total luminosity radiated by the pulsar is then given by $L_\gamma=4\pi f_\Omega F_{obs}D^2$ where $F_{obs}$ is the observed phase-averaged energy flux over $100$ MeV and $D=250_{+120}^{-62}$ pc is the pulsar distance \citep{faherty07}. The estimated averaged luminosity is then $L_{\gamma}$=3.1$\times$10$^{34}f_\Omega$ erg s$^{-1}$, yielding a $\gamma$-ray efficiency $\eta_{\gamma}=\frac{L_\gamma}{\dot{E}}$ = 0.15$f_{\Omega}$ ($d$/100pc)$^{2}$.\\ 
Ideally, geometrical values in Table \ref{table_watters} should be compared with independent estimates, coming e.g. for radio polarization or from the geometry of the pulsar wind nebula \citep{NgRomani04,NgRomani08}.\\
Owing to the lack of radio emission, the only geometrical constraints available for Geminga come from the X-ray observations which have unveiled a faint bow shock structure, due to the pulsar motion in the interstellar medium \citep{caraveo03} and a inner tail structure \citep{deluca06,pavlov06}, while phase resolved spectroscopy yielded a glimpse of the geometry of the emitting regions as the neutron star rotates \citep{caraveo04}.\\
The shape of the bow shock feature constrains its inclination to be less than 30$^{\circ}$ with respect to the plane of the sky. Since such a feature is driven by the neutron star proper motion, the constraint applies also to the pulsar proper motion vector and thus, presumably, to its rotation axis, as is the case for the Vela Pulsar \citep{caraveo01}, pointing to an earth viewing angle ranging from  60 to 90 degrees.\\
Analysing  the pulsar spectral components along its rotational phase, \citet{caraveo04} concluded that the observed behaviour could be explained in the frame of an almost aligned rotator seen at high inclination.\\ 
However rough, such constraints would definitely favour the OG model pointing to a beaming factor of 0.1-0.15. Such a value turns out to be in agreement also with the heuristic luminosity law $\eta\simeq\left(\dot{E}/10^{33}\right)^{-0.5}$ given by \citet{arons96} and \citet{watters09}, that for the Geminga parameters should yield a value of $\sim$ 17$\%$. For the nominal parallax distance of 250 pc, a beaming factor of 0.15 would yield a luminosity of $L_{\gamma}$=4.6$\times$10$^{33}$ erg s$^{-1}$.\\
We note that TPC models, characterized by higher efficiency,  would yield higher luminosity which would account for the
entire rotational energy loss for a distance of  $\sim$300 pc, well within the distance uncertainty. On the other hand, a 100$\%$ efficiency would translate into a distance of 730 pc for the OG model, providing a firm limit on the maximum source distance.
\subsection{Phase resolved spectroscopy}
The power law with exponential cut-off describes only approximately the phase-averaged spectrum of Geminga, since several spectral components contribute at different rotational phases. The phase-resolved analysis that we have performed is thus a powerful tool for probing the emission of the Geminga pulsar.\\
Figure \ref{fig:ph_evol} shows a sudden change in the spectral index around each peak maximum. The spectrum appears to be very hard in the ``first interpeak'' region between P1 and P2,  with an index close to $\Gamma$ $\sim$ 1.1 and softens quickly after the peak maximum and in the "second interpeak" to $\Gamma$ $\sim$ 1.5. Caustic models such as OG and TPC predict such behavior as a result of the change in emission altitude with energy. Sudden changes in the energy cut-off are also predicted, as
is also seen for Geminga. Large variations in the spectral index and energy cut-off as a function of the pulsar phase have already been seen in other pulsars, such as the Crab pulsar \citep{abdo10c} or PSR J2021+3651 \citep{abdo09e}.\\
The persistence of an energy cut-off in the "second interpeak" region suggests pulsar emission extending over the whole rotation, further supporting the TPC model for Geminga.
A similar ``second interpeak'' has been also observed by Fermi-LAT in PSR J1836+5925,
known as the ``next Geminga'' \citep{halpern07}. Although
Geminga is significantly younger, the two pulsars share other interesting features,
including very similar spectral spectral indexes and energy cut-offs in the
phase-averaged spectrum, and comparable X-ray spectra \citep{abdo10d}.
\section{Conclusions}
In this Paper we presented the analysis of Geminga based on data collected during the first year of $Fermi$ operations. The large collecting area of the LAT allows a timing solution to be obtained solely from $\gamma$-ray data.\\ 
The study of the light curve showed the evolution of the pulse profile with energy, unveiling the shrinking of the peaks with increasing energy and providing insights on the highest energies with unprecedented detail. Although the phase-averaged spectrum is consistent with a power law with exponential cut-off, the phase-resolved analysis showed a much richer picture of different spectral components intervening at different rotational phases. The phase-resolved analysis has also allowed the detection of the ``second interpeak'' emission indicating a pulsar emission extending over all phases. 
This feature, never seen before in Geminga, was recently also seen by $Fermi$ LAT in PSR J1836+5925 \citep{abdo10d}.

Our results favor the outer magnetospheric origin for the $\gamma$-ray emission.  The distance uncertainty allows for reasonable values of efficiency for both OG and TPC models, although the efficiency for the TPC model becomes too large for distance values just above the nominal one. Future improvements in estimating the distance of Geminga will help to better strengthen the conclusions and constraining outer magnetospheric models.\\
The light curve and phase-resolved spectral studies provide a much stronger constraint on the model geometry.  The inclination and viewing angle phase space for peak separation of 0.5 is very small for the OG, which however provides values compatible with those obtained from the analysis of Geminga's X-ray behaviour. On the other hand, TPC geometry would seem more natural for pulsars of Geminga's age that have large gaps. Pulsed emission at all phases is a common feature of the TPC geometry. It occurs infrequently for OG geometries, although is present for the large $\zeta_{E}$ solutions invoked here for Geminga.
\acknowledgments
The \textit{Fermi} LAT Collaboration acknowledges generous ongoing support
from a number of agencies and institutes that have supported both the
development and the operation of the LAT as well as scientific data analysis.
These include the National Aeronautics and Space Administration and the
Department of Energy in the United States, the Commissariat \`a l'Energie Atomique
and the Centre National de la Recherche Scientifique / Institut National de Physique
Nucl\'eaire et de Physique des Particules in France, the Agenzia Spaziale Italiana
and the Istituto Nazionale di Fisica Nucleare in Italy, the Ministry of Education,
Culture, Sports, Science and Technology (MEXT), High Energy Accelerator Research
Organization (KEK) and Japan Aerospace Exploration Agency (JAXA) in Japan, and
the K.~A.~Wallenberg Foundation, the Swedish Research Council and the
Swedish National Space Board in Sweden.
Additional support for science analysis during the operations phase is gratefully
acknowledged from the Istituto Nazionale di Astrofisica in Italy and the Centre National d'\'Etudes Spatiales in France.
\section*{Appendix A: detailed results from phase-resolved spectral analysis}
\begin{figure}[ht]
\begin{center}
\includegraphics[width=0.95\columnwidth]{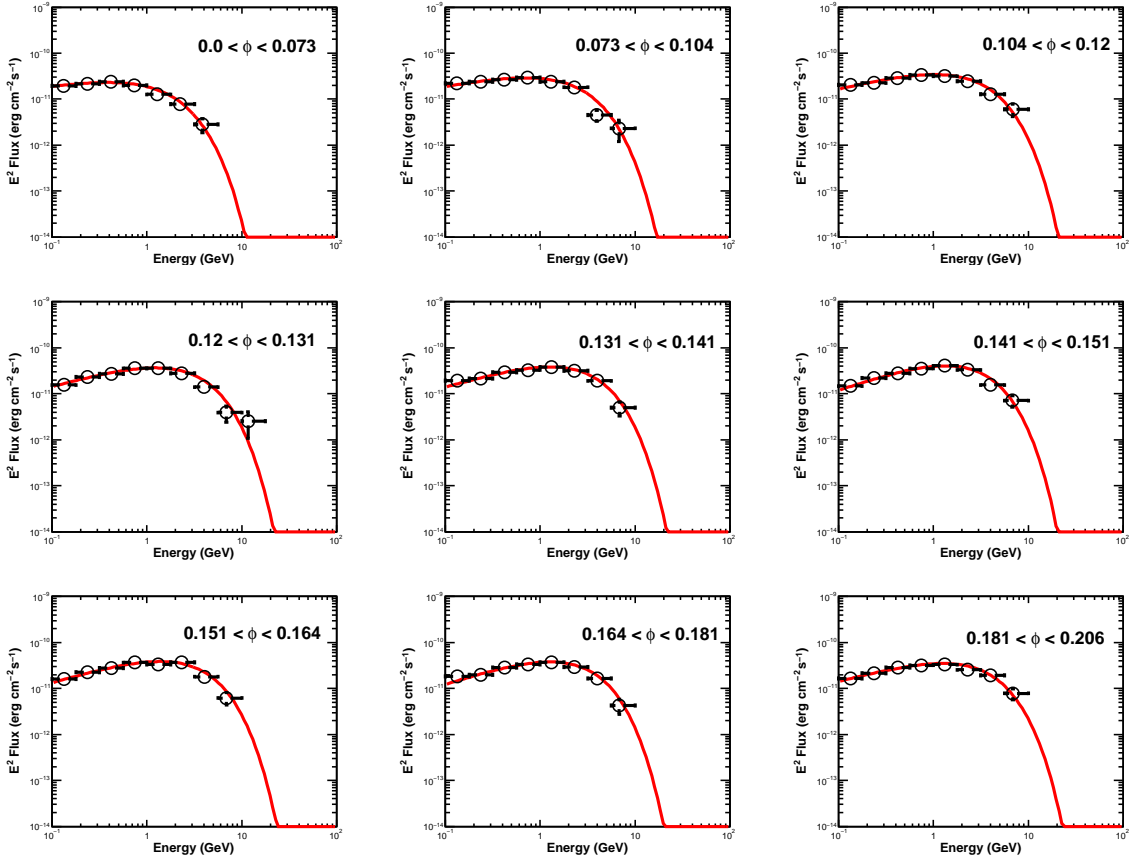}
\caption{Phase-resolved Spectral Energy Distributions (SEDs) of the Geminga pulsar in the phase range $\phi$ = 0.0 - 0.206. The spectral parameters of each of these spectral distributions can be found in Table \ref{tab:phres}. The fluxes are not normalized to the phase bin width, whereas in Table \ref{tab:phres} the fluxes are normalized. The curves represent the best fit power law with exponential cut-off, while the LAT spectral points (open circles) are obtained using the maximum likelihood method described in Section \ref{sec:an_spec}}\label{fig:app_phres1}
\end{center}
\end{figure}

\begin{figure}[ht]
\begin{center}
\includegraphics[width=0.95\columnwidth]{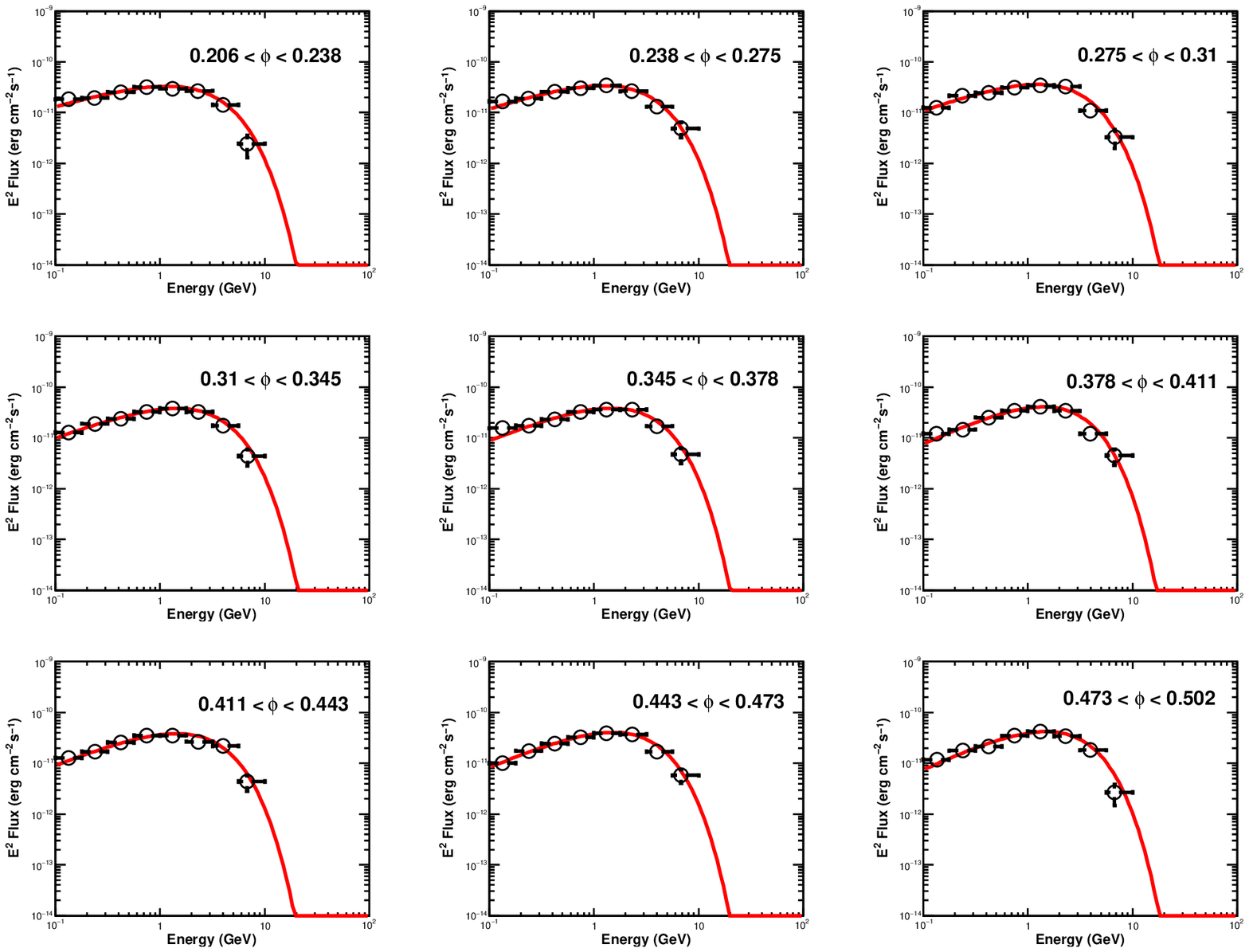}
\caption{Phase-resolved Spectral Energy Distributions (SEDs) of the Geminga pulsar in the phase range $\phi$ = 0.206 - 0.502.}\label{fig:app_phres2}
\end{center}
\end{figure}

\begin{figure}[ht]
\begin{center}
\includegraphics[width=0.95\columnwidth]{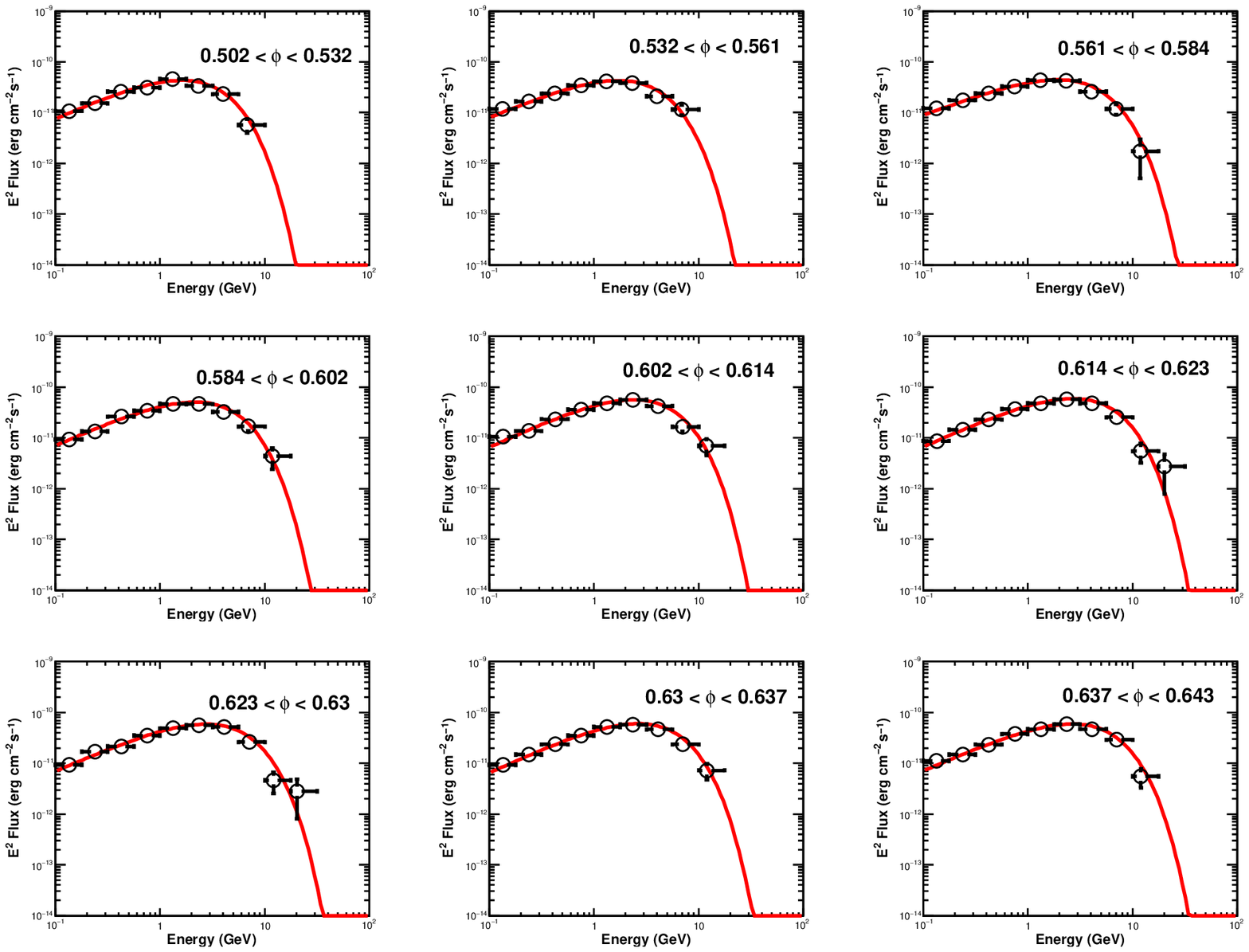}
\caption{Phase-resolved Spectral Energy Distributions (SEDs) of the Geminga pulsar in the phase range $\phi$ = 0.502 - 0.643.}\label{fig:app_phres3}
\end{center}
\end{figure}

\begin{figure}[ht]
\begin{center}
\includegraphics[width=0.95\columnwidth]{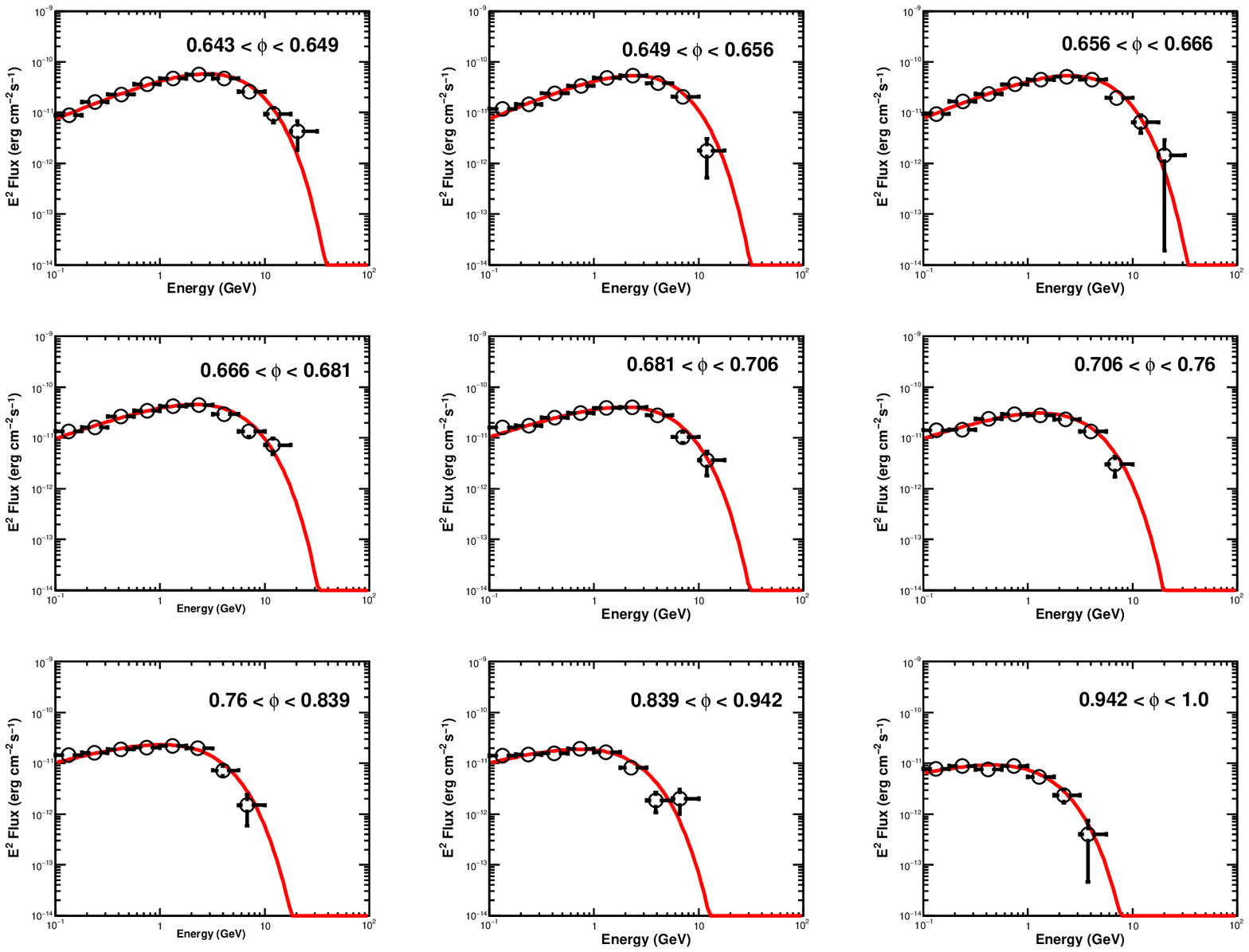}
\caption{Phase-resolved Spectral Energy Distributions (SEDs) of the Geminga pulsar in the phase range $\phi$ = 0.643 - 1.0.}\label{fig:app_phres4}
\end{center}
\end{figure}

In this Appendix we report all the numerical results and the spectral Energy Distributions (SEDs) obtained from the phase-resolved spectral analysis of Geminga. Table \ref{tab:phres} shows the spectral parameters obtained from the spectral fit in each phase interval, while Figures from \ref{fig:app_phres1} to \ref{fig:app_phres4} show the plots of all the SEDs.
\begin{table}[ht]
\begin{center}
\begin{tabular}{||c|c|c|c|c||}
\hline
 $\phi_{min}$ & $\phi_{max}$ & Flux $>$ 0.1 GeV &
Spectral index & Cut-off energy \\
  &  & ($\times$ 10$^{-7}$ cm$^{-2}$ s$^{-1}$) &
 &  (GeV) \\
\hline
0.000  &  0.073  &  1.72  $\pm$  0.07  &  1.67  $\pm$  0.08  &  1.19  $\pm$  0.16   \\                                         
0.073  &  0.104  &  4.49  $\pm$  0.15  &  1.59  $\pm$  0.06  &  1.75  $\pm$  0.21  \\
0.104  &  0.120  &  9.14  $\pm$  0.27  &  1.49  $\pm$  0.05  &  2.04  $\pm$  0.22  \\
0.120  &  0.131  &  12.46  $\pm$  0.36  &  1.40  $\pm$  0.05  &  2.07  $\pm$  0.23  \\
0.131  &  0.141  &  14.24  $\pm$  0.40  &  1.37  $\pm$  0.05  &  2.02  $\pm$  0.19  \\
0.141  &  0.151  &  13.09  $\pm$  0.37  &  1.26  $\pm$  0.05  &  1.84  $\pm$  0.16  \\
0.151  &  0.164  &  10.74  $\pm$  0.31  &  1.37  $\pm$  0.05  &  2.19  $\pm$  0.21  \\
0.164  &  0.181  &  7.76  $\pm$  0.23  &  1.30  $\pm$  0.05  &  1.85  $\pm$  0.17  \\
0.181  &  0.206  &  5.35  $\pm$  0.17  &  1.43  $\pm$  0.05  &  2.21  $\pm$  0.23  \\
0.206  &  0.238  &  3.89  $\pm$  0.13  &  1.39  $\pm$  0.05  &  1.91  $\pm$  0.20  \\
0.238  &  0.275  &  3.30  $\pm$  0.11  &  1.33  $\pm$  0.06  &  1.83  $\pm$  0.18  \\
0.275  &  0.310  &  3.36  $\pm$  0.11  &  1.24  $\pm$  0.06  &  1.64  $\pm$  0.15  \\
0.310  &  0.345  &  3.29  $\pm$  0.11  &  1.22  $\pm$  0.05  &  1.86  $\pm$  0.17  \\
0.345  &  0.378  &  3.36  $\pm$  0.11  &  1.16  $\pm$  0.06  &  1.76  $\pm$  0.16  \\
0.378  &  0.411  &  3.24  $\pm$  0.11  &  1.02  $\pm$  0.06  &  1.43  $\pm$  0.12  \\
0.411  &  0.443  &  3.51  $\pm$  0.11  &  1.16  $\pm$  0.06  &  1.70  $\pm$  0.15  \\
0.443  &  0.473  &  3.70  $\pm$  0.12  &  1.11  $\pm$  0.06  &  1.72  $\pm$  0.15  \\
0.473  &  0.502  &  3.63  $\pm$  0.12  &  1.02  $\pm$  0.06  &  1.53  $\pm$  0.12  \\
0.502  &  0.532  &  3.64  $\pm$  0.12  &  1.04  $\pm$  0.06  &  1.68  $\pm$  0.14  \\
0.532  &  0.561  &  3.82  $\pm$  0.12  &  1.10  $\pm$  0.05  &  1.91  $\pm$  0.16  \\
0.561  &  0.584  &  4.78  $\pm$  0.15  &  1.19  $\pm$  0.05  &  2.38  $\pm$  0.21  \\
0.584  &  0.602  &  6.21  $\pm$  0.18  &  1.06  $\pm$  0.05  &  2.31  $\pm$  0.18  \\
0.602  &  0.614  &  9.26  $\pm$  0.26  &  1.02  $\pm$  0.04  &  2.47  $\pm$  0.18  \\
0.614  &  0.623  &  12.67  $\pm$  0.35  &  1.05  $\pm$  0.04  &  2.85  $\pm$  0.22  \\
0.623  &  0.630  &  15.16  $\pm$  0.41  &  1.09  $\pm$  0.04  &  3.01  $\pm$  0.24  \\
0.630  &  0.637  &  16.50  $\pm$  0.44  &  1.03  $\pm$  0.04  &  2.73  $\pm$  0.20  \\
0.637  &  0.643  &  17.78  $\pm$  0.48  &  1.08  $\pm$  0.04  &  2.88  $\pm$  0.22 \\
0.643  &  0.649  &  17.88  $\pm$  0.48  &  1.13  $\pm$  0.04  &  3.24  $\pm$  0.26  \\
0.649  &  0.656  &  15.89  $\pm$  0.44  &  1.10  $\pm$  0.04  &  2.66  $\pm$  0.21  \\
0.656  &  0.666  &  11.74  $\pm$  0.33  &  1.13  $\pm$  0.04  &  2.89  $\pm$  0.23  \\
0.666  &  0.681  &  8.14  $\pm$  0.24  &  1.26  $\pm$  0.04  &  2.91  $\pm$  0.26  \\
0.681  &  0.706  &  4.67  $\pm$  0.15  &  1.32  $\pm$  0.05  &  2.83  $\pm$  0.30  \\
0.706  &  0.760  &  1.94  $\pm$  0.07  &  1.30  $\pm$  0.06  &  1.86  $\pm$  0.20  \\
0.760  &  0.839  &  1.18  $\pm$  0.05  &  1.42  $\pm$  0.07  &  1.79  $\pm$  0.23  \\
0.839  &  0.942  &  0.83  $\pm$  0.04  &  1.44  $\pm$  0.09  &  1.30  $\pm$  0.19  \\
0.942  &  1.000  &  0.81  $\pm$  0.06  &  1.48  $\pm$  0.17  &  0.87  $\pm$  0.19  \\
\hline
\end{tabular}
\end{center}
\caption{Phase interval definitions and corresponding spectral parameters obtained from fitting the spectrum with a power law with exponential cut-off. The flux in the third column is normalized to the width of the phase bin. The systematic uncertainties are in agreement with the ones evaluated for the phase averaged analysis.}
\label{tab:phres}
\end{table}

\end{document}